  \documentstyle[preprint,aps]{revtex}

\newcommand{\ppb}{\overline{p}p}
\newcommand{\NNb}{\overline{N}N}

\begin{document}
\preprint{KFA-IKP(TH)-1994-42}
\draft
%===================================================================
\title
{\bf
ROLE OF MESON-MESON CORRELATION EFFECTS \\
IN THE $\bf\NNb\to\rho\pi$ ANNIHILATION PROCESS
}
%===================================================================
\author
{V. Mull, G. Janssen and J. Speth}
%===================================================================
\address
{Institut f\"ur Kernphysik, Forschungszentrum J\"ulich GmbH,
D-52425 J\"ulich, Germany}
%===================================================================
\author
{K. Holinde}
%===================================================================
\address
{Departamento de Fisica Teorica, Facultad de Ciencias Fisicas,
Universidad de Valencia, Burjasot (Valencia), Spain \\
and \\
Institut f\"ur Kernphysik, Forschungszentrum J\"ulich GmbH,
D-52425 J\"ulich, Germany}
%===================================================================
\date{\today}

\maketitle
\begin{abstract}
Meson-meson correlation effects are investigated in the
$\NNb\to\rho\pi$ annihilation process using a realistic meson-exchange model
for the $\rho\pi$ interaction determined previously, together with a
conventional baryon-exchange transition model and a consistent $\NNb$
interaction. For $\NNb$ $S$-states, they have a drastic effect and bring
the relative (${^1S_0}/{^3S_1}$) branching ratio up to the experimental
value, thus resolving the long-standing so-called ``$\rho\pi$'' puzzle.
For $\NNb$ $P$-states, their effect is of minor importance,
and discrepancies remain for those ratios involving annihilation from the
$\NNb({^3P_J})$ state to $\rho\pi(l'=2)$.
\end{abstract}
%===================================================================

Reactions involving antinucleons (for a review see e.g.\ the papers by Amsler
and Myhrer \cite{AM} and Dover et al. \cite{DGMF}) have always been considered
to be the ideal place for finding quark effects since annihilation phenomena
from the antinucleon-nucleon ($\NNb$) system are supposed to be governed by
short distance physics. However, if one tries to discriminate between different
theoretical scenarios for the microscopic description of the annihilation
process, it is not sufficient to look at global features like e.g.\ the total
annihilation cross section.  Rather, one should deal with specific annihilation
channels and consider experimental information for annihilation from specific
$\NNb$ initial states.  Indeed, corresponding experimental results from LEAR
for
two-meson annihilation (for references see \cite{DGMF}) are of particular
interest since they provide clear evidence for the dynamical suppression of
transitions which are in principle allowed by the conservation of fundamental
quantum numbers.  Such dynamical selection rules yield valuable information
about the transition process and should impose in principle severe restrictions
on the theoretical description of the annihilation mechanism, e.g.\ in terms of
conventional baryon exchange or explicit quark-gluon exchange.  Unfortunately,
things are very much obscured by the presence of initial ($\NNb$) and final
(meson-meson) state interactions.  Although it was often argued that the
consideration of relative ratios should minimize these effects it turned out
(see e.g.\ \cite{NNbII,NNbIII}) that there is a strong sensitivity to whether
and even which kind of initial state interaction is included. The latter does
not drop out even if ratios from the same partial wave are considered. The
conclusion drawn in Refs.\cite{NNbII,NNbIII} was that a consistent description
for the transition model and initial state $\NNb$ interaction is required
before
we can seriously address the question about which transition mechanism is
preferred.  Here we will demonstrate that also, at least in some specific
channels, meson-meson correlation effects have a considerable influence
on the explanation of the experimental data.

The most prominent example for the realization of such a dynamical selection
rule is the so-called ``$\rho\pi$'' puzzle in the $\NNb\to\rho\pi$ annihilation
process. Already in the sixties, $\rho\pi$ branching ratios in liquid hydrogen
have been determined \cite{Foster}, the results being essentially the same for
all charge combinations $\rho^+\pi^-, \rho^-\pi^+, \rho^0\pi^0$. The
$\rho^0\pi^0$ combination, being a pure isospin-zero state, can be produced
only
from the (${^3S_1},I=0$) protonium state whereas the charged combinations can
also be generated from the (${^1S_0},I=1$) state.  (Production from the
(${^3S_1},I=1$) and (${^1S_0},I=0$) states is strictly forbidden due to
$G$-parity
conservation.)  Since all $S$-states in protonium should be populated with
about
the same probability the annihilation from the ${^1S_0}$ state in the $\rho\pi$
channel is obviously strongly suppressed.  In $P$-waves, however, such a
supression of annihilation from $I=1$ states cannot be seen in the recent
results of
the ASTERIX experiment\cite{May}.

In Table \ref{tabI}, present empirical information about these ratios is
compared with theoretical results, which we obtained recently
\cite{NNbII,NNbIII} using a model for the transition process based on baryon
($N,\Delta$) exchange, see Fig.\ \ref{figI}. An initial state interaction is
either completely neglected (Born) or included (A(BOX)\cite{NNbI},
D\cite{NNbIII}). Both $\NNb$ interactions use as elastic part the $G$-parity
transformed (full) Bonn $NN$ potential \cite{MHE}. In model A(BOX),
annihilation
has been accounted for by a simple phenomenological, energy and state
independent optical potential with both real and imaginary part and three
adjustable parameters. In model D \cite{NNbIII}, the annihilation part of the
$\NNb$ interaction is split into two parts: Intermediate states with two mesons
are described microscopically in terms of baryon-exchange processes, including
all possible combinations of
$\pi,\eta,\rho,\omega,a_0,f_0,a_1,f_1,a_2,f_2,K,K^*$.  Their strength has been
adjusted in a consistent description of $\NNb$ scattering and annihilation to
the empirical information about the annihilation into specific channels.  The
remaining part (three-meson channels etc.)  is taken into account by a
phenomenological optical potential of similar form as used in model A(BOX)
\cite{NNbI}.

The results of Table \ref{tabI} indeed confirm that there is a considerable
sensitivity to which kind of initial state interaction is used. Still, for the
$S$-state ratio, all theoretical results quoted there are so far off the
experimental values that one is tempted to conclude that the baryon exchange
transition model cannot be appropriate to account for the empirical situation.
There are numerous calculations based on alternative transition models in the
literature \cite{Maruyama,Mundigl,GreNis,Dover}; some of them do give at least
a
better description of the ${^1S_0}/{^3S_1}$ ratio. For example, the model of
Maruyama et al.\cite{Maruyama} using the A2 mechanism (two $\overline{q}q$ pair
annihilations followed by one $\overline{q}q$ pair creation in a planar
topology) and the ${^3P_0}$ vertex, achieves a value of 1:10-18. Thus it
appears
that this quark model might be superior in comparison to conventional
baryon-exchange, at least in this sector.  However, this conclusion is
definitely premature, for the following reason: So far, correlation effects in
the outgoing meson-meson channel have been neglected completely; their
inclusion
in the description of the transition process might well lead to different
theoretical values in Table \ref{tabI}.

In this paper, we want to investigate the role of the final state interaction
in
the $\NNb\to\rho\pi$ transition process; especially we want to see whether its
inclusion will bring the theoretical baryon-exchange model results in better
agreement with the empirical situation.

Unfortunately, not much is known empirically about the interaction between a
pion and a rho-meson since due to the short lifetime of the $\rho$-meson no
$\rho\pi$ scattering data exists. Therefore, one has essentially  to rely on a
dynamical model. Recently we have constructed such a $\rho\pi\to\rho\pi$
amplitude\cite{PiRhoWW} based on both $s$- and $t$-channel meson exchange
diagrams, see Fig.\ \ref{figII}, acting as driving terms in a scattering
equation.
Parameters (coupling constants and cutoff masses in formfactors) have been
partly taken from other investigations, partly adjusted to empirical
information
about $a_1, \omega \to \rho\pi$ decay. The model has been successfully tested
in
the $NN$ system \cite{NNpirho}: By using it as the basic ingredient in the
correlated $\rho\pi$ exchange piece of the $NN$ interaction, it was possible to
obtain sufficient $NN$ tensor force despite of having a soft $\pi NN$
formfactor
demanded by numerous independent information.

The $\rho\pi$ interaction (with all parameters prefixed) is now considered in
our
calculation in two ways: First we include it in a DWBA-type approach in the
final state only, cp.\ Fig.\ \ref{figIII}(a,b). In a second step, we perform a
$\NNb,\rho\pi$ coupled channels calculation; in this way, it is also included
in
the initial state, see Fig.\ \ref{figIII}(c).  The $\NNb$ interaction consists
of course of an elastic and an annihilation part. In order to avoid double
counting the effects of the $\rho\pi$ channel have to be removed from the
$\NNb$
annihilation part in the coupled channels calculation.  Since this can only be
done in a well defined way for the consistent microscopic annihilation model D,
only this model is used in the following.

We start by demonstrating the effect of the $\rho\pi$ interaction on the
$\ppb\to\rho\pi$ cross sections in flight. The inclusion of the meson-meson
interaction leads to a noticeable increase as demonstrated in Fig.\
\ref{figIV}.
On the other hand the new results still lie in the experimental region, so a
readjustment of formfactor parameters in the baryon exchange transition was not
required. In the coupled channel approach, the inclusion of the $\rho\pi$
interaction modifies also the initial $\NNb$ interaction, cp. Fig.\
\ref{figIII}(c). Fortunately, the resulting $\NNb\to\NNb$ observables are
barely
changed since the $\NNb\to\rho\pi$ transition is anyhow a small contribution to
the total annihilation. Thus a readjustment of parameters in this sector is
likewise not necessary.

Table \ref{tabII} contains the resulting branching ratios at rest, for the
$\rho^+\pi^-+\rho^-\pi^+$ and $\rho^0\pi^0$ annihilation channels.  As expected
already from Fig.\ \ref{figIV}, the consideration of the $\rho\pi$ interaction
increases both branching ratios. Its inclusion in both the initial and final
state (CC) leads to a result in good agreement with experiment\cite{AF}.

Let us now look again at the relative branching ratios from specific $\NNb$
partial wave states (cp.\ Table \ref{tabIII}).  The inclusion of the
meson-meson
interaction changes the ${^1S_0}/{^3S_1}$ ratio (corresponding to
$\pi$/$\omega$
quantum numbers respectively) drastically: The change is so large that, in a
DWBA calculation, the result nearly coincides with the empirical value; the
full
coupled channel calculation (which includes the $\rho\pi$ interaction also in
the $\NNb$ annihilation potential, cp.\ Fig.\ \ref{figIII}(c)) leads to a
somewhat smaller ratio well within the experimental error bars.  On the other
hand, the modification of the $P$-state ratios is quite small.  While the
${^1P_1}/{^3P_1(l'=0)}$ ratio is in agreement with experiment, annihilation
from
the $\NNb({^3P})$ state into the $\rho\pi$ system  with relative orbital
angular momentum $l'=2$
provides the largest contribution of all $P$-states, in sharp
contrast to experiment where annihilation from this state appears to be
negligible.

In this context the question arises whether the introduction of the $\rho\pi$
interaction in the $\NNb$ $P$-states keeps not only the ratios essentially the
same but also the absolute values. This is indeed the case: The change of the
absolute $P$-state values is likewise of the order of $10 - 20 \%$ only, as for
the ratios.

As shown in Table \ref{tabIV} the $\rho\pi$ interaction reduces the ${^1S_0}$
contribution strongly while it increases the ${^3S_1}$ contribution leading to
the extremely small ratio of Table \ref{tabIII}, in agreement with
experiment. Note that this drastically different role of the $\rho\pi$
interaction in the ${^1S_0}$ versus ${^3S_1}$ state does not imply a comparably
strong state dependence of this interaction. (In fact, the difference between
the $\rho\pi$ interactions in these partial waves is quite small). If, for
example, we artificially use exactly the same $\rho\pi$ interaction, which
enters the $\ppb$ ${^1S_0}$ state, in both ${^1S_0}$ and ${^3S_1}$, we obtain
qualitatively the same results as before. This fact demonstrates once more that
correlation effects do not drop out even when relative ratios from the same
partial wave are considered.

In summary, the inclusion of the interaction between a pion and a rho in the
$\NNb\to\rho\pi$ annihilation process provides drastic changes in the relative
branching ratios. For $S$-states, it brings the result in agreement with
experiment and is thus a promising candidate to resolve the long-standing
``$\rho\pi$ puzzle''.  Our model agrees with the experimental information also
in case of the ${^1P_1}/{^3P_1(l'=0)}$ ratio, but it is in contrast to the
present
empirical values for ratios involving annihilations from the ${^3P}$ states
into $\rho\pi$ with $l'=2$.  Whether an
extension of our $\rho\pi$ interaction model (e.g.\ by adding a suitable pole
term in the relevant partial wave or considering the coupling to additional
channels, for example $\overline{K}K^*\pm\overline{K^*}K$), or improved
data can solve this problem remains to be seen. In any
case and in more general terms, meson-meson correlation effects have to be
included and treated in a consistent way before one can seriously address the
question about the relevant transition mechanism in $\NNb$ annihilation.

Discussions with E.\ Klempt concerning the interpretation of the
ASTERIX data \cite{May} are gratefully acknowledged.
One of the authors (K.H.) would like to thank the Direcci\'on General de
Investigacion Cientifica y Tecnica (DGICYT) for financial support making his
stay in Valencia possible (SAB94-0218), and the Department of Theoretical
Physics, especially Prof.\ Oset for the generous hospitality.

%%%%%%%%%%%%%%%%%%%%%%%%%%%%%%%%%%%%%%%%%%%%%%%%%%%%%%%%%%%%%%%%%%%%%%%%%
%%%%%{Table I}
\begin{table}% [t]
%\squeezetable
\caption{
\label{tabI}
Ratios of branching ratios ``at rest'' for the annihilation
$\ppb\to\rho^\pm\pi^\mp$.  The experimental values are calculated from data
given in Ref.\protect\cite{May}.  The theoretical results are obtained as
relative cross sections at $p_{lab}=100 MeV/c$.  The theoretical values are
based on a transition model pictorially described in Fig.\ \protect\ref{figI},
i.e.\ without final state interaction. The initial state interaction is either
neglected, too (Born) or taken from model A(BOX)\protect\cite{NNbI} and D
\protect\cite{NNbIII}.  }
%%%%%%%%%%%%%%%%%%%%%%%%%%%%%%%%%%%%%%%%%%%%%%%%%%%%%%%%%%%%%%%%%%%%%%%%%
\tiny
\begin{tabular}{c|ccc |c |c}
                             &Born  &A(BOX)&  D    & Exp. \cr
%%%%%%%%%%%%%%%%%%%%%%%%%%%%%%%%%%%%%%%%%%%%%%%%%%%%%%%%%%%%%%
\noalign{\hrule}
${\ppb ({^1S_0},I=1)}$ \ :   & 1    & 1    & 1     &  1            \cr
${\ppb ({^3S_1},I=0)}$       & 3.45 & 2.29 & 4.67  &  35$\pm$16    \cr
\noalign{\hrule}
${\ppb ({^1P_1},I=0)}$\ :    & 1    & 1    & 1     &  1            \cr
${\ppb ({^3P_{1,2}},I=1)}$   & 5.28 & 9.10 & 2.95  &  0.84$\pm$0.22  \cr
\end{tabular}
\end{table}
%%%%%%%%%%%%%%%%%%%%%%%%%%%%%%%%%%%%%%%%%%%%%%%%%%%%%%%%%%%%%%%%%%%%%%%%%

%%%%%%%%%%%%%%%%%%%%%%%%%%%%%%%%%%%%%%%%%%%%%%%%%%%%%%%%%%%%%%%%%%%%%%%%%
%%%%%{table II}
\begin{table}%[ht]

\caption{
\label{tabII}
Branching ratios ``at rest'' for annihilation into $\rho\pi$.  The data are
taken from Ref.\protect\cite{AF}.  The theoretical results are obtained as
relative cross sections at $p_{lab}=100 MeV/c$.  The first column denotes the
results based on the transition model without final state interaction (Fig.\
\protect\ref{figI}) using model D \protect\cite{NNbIII} as initial state
interaction. In the second (third) column, the meson-meson interaction is
included in a DWBA calculation (as final state interaction) or in a full
$\NNb,\rho\pi$ coupled-channel (CC) calculation.  }
%%%%%%%%%%%%%%%%%%%%%%%%%%%%%%%%%%%%%%%%%%%%%%%%%%%%%%%%%%%%%%%%%%%%%%%%
\begin{tabular}{c|c|c|c|c}
%%%%%%%%%%%%%%%%%%%%%%%%%%%%%%%%%%%%%%%%%%%%%%%%%%%%%%%%%%%%%%%%%%%%%%%%
$\overline{p} p  \to $      & no FSI & DWBA   & CC & EXP.\  \cr
\noalign{\hrule}
%%%%%%%%%%%%%%%%%%%%%%%%%%%%%%%%%%%%%%%%%%%%%%%%%%%%%%%%%%%%%%%%%%%%%%%%
$
\rho^{+}\pi^{-}+\rho^{-}\pi^{+}$
 & 2.32   & 2.50   &2.73    &$ 3.4  \pm0.2 $ \cr
$\rho^{\scriptscriptstyle 0} \pi^{\scriptscriptstyle 0} $
 & 0.85   & 1.08   &1.19    &$ 1.4  \pm0.1 $ \cr
\end{tabular}
\end{table}
%%%%%%%%%%%%%%%%%%%%%%%%%%%%%%%%%%%%%%%%%%%%%%%%%%%%%%%%%%%%%%%%%%%%%%%%%
%%%%%{Table III}
\begin{table}%[t]
\ifpreprintsty
\else
\squeezetable
\fi
\caption{
\label{tabIII}
Ratios of branching ratios ``at rest'' for the annihilation
$\ppb\to\rho^\pm\pi^\mp$.  The experimental values are calculated from data
given in Ref.\protect\cite{May}.  The theoretical results are obtained as
relative cross sections at $p_{lab}=100 MeV/c$.  The first column denotes the
results based on the transition model without final state interaction (Fig.\
\protect\ref{figI}) using model D \protect\cite{NNbIII} as initial state
interaction. In the second (third) column, the meson-meson interaction is
included in a DWBA calculation as final state interaction or in a full
$\NNb,\rho\pi$ coupled-channels (CC) calculation.
$l'$ denotes the orbital angular momentum of the $\rho\pi$ system.}
%%%%%%%%%%%%%%%%%%%%%%%%%%%%%%%%%%%%%%%%%%%%%%%%%%%%%%%%%%%%%%%%%%%%%%%%%
%\small
\begin{tabular}{c|c|c|c|c}
            &no FSI & DWBA & CC & Exp.\ \cr
%%%%%%%%%%%%%%%%%%%%%%%%%%%%%%%%%%%%%%%%%%%%%%%%%%%%%%%%%%%%%%%%%%%%%%%%
\noalign{\hrule}
${\ppb ({^1S_0},I=1)}$\ :\ & 1    & 1    & 1    & 1             \cr
${\ppb ({^3S_1},I=0)}$     & 4.67 & 32.4 & 23.9 & 35$\pm$16     \cr
\noalign{\hrule}
${\ppb ({^1P_1},I=0)}$\ :\ & 1    & 1    & 1    & 1             \cr
${\ppb ({^3P_{1,2}},I=1)}$ & 2.95 & 3.02 & 2.95 & 0.84 $\pm$0.22  \cr
\noalign{\hrule}
${\ppb ({^1P_1},I=0)}$\ :\ & 1    & 1    & 1    & 1             \cr
${\ppb ({^3P_1})\to l'=0}$ & 0.54 & 0.68 &0.64  & 0.8 $\pm$0.2  \cr
\noalign{\hrule}
${\ppb ({^1P_1},I=0)}$\ :\ & 1    & 1    & 1    & 1             \cr
${\ppb ({^3P_{1,2}})\to l'=2}$
                           & 2.41 &2.34  &2.31  & 0.04 $\pm$0.015  \cr
$({^3P_{1}}+{^3P_{2}})$ &(0.36+2.05)&(0.20+2.14)&(0.20+2.11)&          \cr
%%%%%%%%%%%%%%%%%%%%%%%%%%%%%%%%%%%%%%%%%%%%%%%%%%%%%%%%%%%%%%%%%%%%%%%%%
\end{tabular}
\end{table}
%%%%%%%%%%%%%%%%%%%%%%%%%%%%%%%%%%%%%%%%%%%%%%%%%%%%%%%%%%%%%%%%%%%%%%%%%

%%%%%%%%%%%%%%%%%%%%%%%%%%%%%%%%%%%%%%%%%%%%%%%%%%%%%%%%%%%%%%%%%%%%%%%%%
%%%%%{Table IV}
\begin{table}%[t]
%%\squeezetable
\caption{
\label{tabIV}
 Partial cross sections in $[\mu b]$ at $p_{lab}=100 MeV/c$ for
 the annihilation $\ppb\to\rho^\pm\pi^\mp$ from $\NNb$ $S$-states.  The first
 column denotes the results based on the transition model without final state
 interaction (Fig.\ \protect\ref{figI}) using model D \protect\cite{NNbIII} as
 initial state interaction. In the second (third) column, the meson-meson
 interaction is included in a DWBA calculation as final state interaction or
 in a full $\NNb,\rho\pi$ coupled-channels (CC) calculation.}
%%%%%%%%%%%%%%%%%%%%%%%%%%%%%%%%%%%%%%%%%%%%%%%%%%%%%%%%%%%%%%%%%%%%%%%%%

\begin{tabular}{c|c|c|c}
            &no FSI & DWBA & CC \cr
%%%%%%%%%%%%%%%%%%%%%%%%%%%%%%%%%%%%%%%%%%%%%%%%%%%%%%%%%%%%%%%%%%%%%%%%
\noalign{\hrule}
${\ppb ({^1S_0},I=1)}$
                         & 0.727      & 0.137      & 0.203    \cr
${\ppb ({^3S_1},I=0)}$
                         & 3.396      & 4.423      & 4.854    \cr
%${\ppb ({^3P_1},I=1)}$
%                         & 0.183     & 0.167      & 0.162    \cr
%${\ppb ({^1P_1},I=0)}$
%                         & 0.233     & 0.249      & 0.252  \cr
%${\ppb ({^3P_2},I=0)}$
%                         & 0.417     & 0.403      & 0.405  \cr
%%%%%%%%%%%%%%%%%%%%%%%%%%%%%%%%%%%%%%%%%%%%%%%%%%%%%%%%%%%%%%%%%%%%%%%%%
\end{tabular}
\end{table}
%%%%%%%%%%%%%%%%%%%%%%%%%%%%%%%%%%%%%%%%%%%%%%%%%%%%%%%%%%%%%%%%%%%%%%%%%

%%%\pagebreak

%%%%%%%%%%%%%%%%%%%%%%%%%%%%%%%%%%%%%%%%%%%%%%%%%%%%%%%%%%%%%%%%%%%%%%%%%
%%%%%{Figure 1}
\begin{figure}
%\vskip 5.5cm
%\special{psfile=br_f1.ps    hoffset=-010 voffset=-050 hscale=050 vscale=050}
\caption{
\label{figI}
$\NNb\to\rho\pi$ transition model without final state interaction as used
in \protect\cite{NNbII,NNbIII}.
}
\end{figure}
%%%%%%%%%%%%%%%%%%%%%%%%%%%%%%%%%%%%%%%%%%%%%%%%%%%%%%%%%%%%%%%%%%%%%%%%%

%%%%%%%%%%%%%%%%%%%%%%%%%%%%%%%%%%%%%%%%%%%%%%%%%%%%%%%%%%%%%%%%%%%%%%%%%
%%%%%{Figure 2}
\begin{figure}
\caption{
\label{figII}
Driving terms used to construct our $\rho\pi\to\rho\pi$ amplitude.
}
\end{figure}
%%%%%%%%%%%%%%%%%%%%%%%%%%%%%%%%%%%%%%%%%%%%%%%%%%%%%%%%%%%%%%%%%%%%%%%%%

%%%%%%%%%%%%%%%%%%%%%%%%%%%%%%%%%%%%%%%%%%%%%%%%%%%%%%%%%%%%%%%%%%%%%%%%%
%%%%%{Figure 3}
\begin{figure}
\caption{
\label{figIII}
$\NNb\to\rho\pi$ transition model of the present paper including the
$\rho\pi$ interaction (a). In the DWBA-type approach based on the
$\NNb$ interaction model D\protect\cite{NNbIII}, processes (b) are
included whereas (c) is not. In an alternative full coupled channels
(CC) calculation, the latter is also taken into account.
}
\end{figure}
%%%%%%%%%%%%%%%%%%%%%%%%%%%%%%%%%%%%%%%%%%%%%%%%%%%%%%%%%%%%%%%%%%%%%%%%%

%%%%%%%%%%%%%%%%%%%%%%%%%%%%%%%%%%%%%%%%%%%%%%%%%%%%%%%%%%%%%%%%%%%%%%%%%
%%%%%{Figure 4}
\begin{figure}
\caption{
\label{figIV}
$\NNb\to\rho\pi$ cross sections in flight. The data are taken from
\protect\cite{Sai}.  For the dash-dotted (dashed) curve, the $\rho\pi$
interaction is neglected (included) and the initial state $\NNb$ interaction
model is taken to be model D \protect\cite{NNbIII}.  The $\NNb,\rho\pi$ coupled
channels approach including the $\rho\pi$ interaction yields the solid line.  }
\end{figure}
%%%%%%%%%%%%%%%%%%%%%%%%%%%%%%%%%%%%%%%%%%%%%%%%%%%%%%%%%%%%%%%%%%%%%%%%%
%
%%%%%%%%%%%%%%%%%%%%%%%%%%%%%%%%%%%%%%%%%%%%%%%%%%%%%%%%%%%%%%%%%%%%%%%%%%
%%%%%%{Figure 5}
%\begin{figure}
%\caption{
%\label{figV}
%}
%\end{figure}
%%%%%%%%%%%%%%%%%%%%%%%%%%%%%%%%%%%%%%%%%%%%%%%%%%%%%%%%%%%%%%%%%%%%%%%%%

\end{document}